\documentclass[acmsmall]{acmart}
\AtBeginDocument{%
  }

\setcopyright{rightsretained}
\acmDOI{10.1145/3698136}
\acmYear{2024}
\acmJournal{PACMHCI}
\acmVolume{8}
\acmNumber{ISS}
\acmArticle{536}
\acmMonth{12}
\received{2024-07-01}
\received[accepted]{2024-09-20}

\begin{document}

\title{Investigating Creation Perspectives and Icon Placement Preferences for On-Body Menus in Virtual Reality}

\author{Xiang Li}
\orcid{0000-0001-5529-071X}
\affiliation{%
  \institution{University of Cambridge}
  \city{Cambridge}
  \country{United Kingdom}}
\email{xl529@cam.ac.uk}

\author{Wei He}
\orcid{0009-0001-6745-2752}
\affiliation{%
  \institution{The Hong Kong University of Science and Technology (Guangzhou)}
  \city{Guangzhou}
  \country{China}}
\email{danielweihe@hkust-gz.edu.cn}

\author{Shan Jin}
\orcid{0000-0001-8676-4033}
\affiliation{%
  \institution{The Hong Kong University of Science and Technology (Guangzhou)}
  \city{Guangzhou}
  \country{China}}
\email{sjin752@connect.hkust-gz.edu.cn}

\author{Jan Gugenheimer}
\orcid{0000-0002-6466-3845}
\affiliation{%
 \institution{TU Darmstadt}
 \city{Darmstadt}
 \country{Germany}}
\affiliation{
  \institution{LTCI, Telecom Paris, IP Paris}
  \city{Palaiseau}
  \country{France}}
\email{jan.gugenheimer@tu-darmstadt.de}
 
\author{Pan Hui}
\orcid{0000-0001-6026-1083}
\affiliation{%
  \institution{The Hong Kong University of Science and Technology (Guangzhou)}
  \city{Guangzhou}
  \country{China}}
\email{panhui@ust.hk}

\author{Hai-Ning Liang}
\orcid{0000-0003-3600-8955}
\affiliation{%
  \institution{The Hong Kong University of Science and Technology (Guangzhou)}
  \city{Guangzhou}
  \country{China}}
\email{hainingliang@hkust-gz.edu.cn}

\author{Per Ola Kristensson}
\orcid{0000-0002-7139-871X}
\affiliation{%
  \institution{University of Cambridge}
  \city{Cambridge}
  \country{United Kingdom}}
\email{pok21@cam.ac.uk}

\renewcommand{\shortauthors}{Xiang Li, et al.}

\begin{abstract}
On-body menus present a novel interaction paradigm within Virtual Reality (VR) environments by embedding virtual interfaces directly onto the user's body. Unlike traditional screen-based interfaces, on-body menus enable users to interact with virtual options or icons visually attached to their physical form. In this paper, We investigated the impact of the creation process on the effectiveness of on-body menus, comparing first-person, third-person, and mirror perspectives. Our first study ($N$ = 12) revealed that the mirror perspective led to faster creation times and more accurate recall compared to the other two perspectives. To further explore user preferences, we conducted a second study ($N$ = 18) utilizing a VR system with integrated body tracking. By combining distributions of icons from both studies ($N$ = 30), we confirmed significant preferences in on-body menu placement based on icon category (\textit{e.g.}, Social Media icons were consistently placed on forearms). We also discovered associations between categories, such as Leisure and Social Media icons frequently co-occurring. Our findings highlight the importance of the creation process, uncover user preferences for on-body menu organization, and provide insights to guide the development of intuitive and effective on-body interactions within virtual environments.
\end{abstract}


\begin{teaserfigure}
  \centering
  \includegraphics[width=\linewidth]{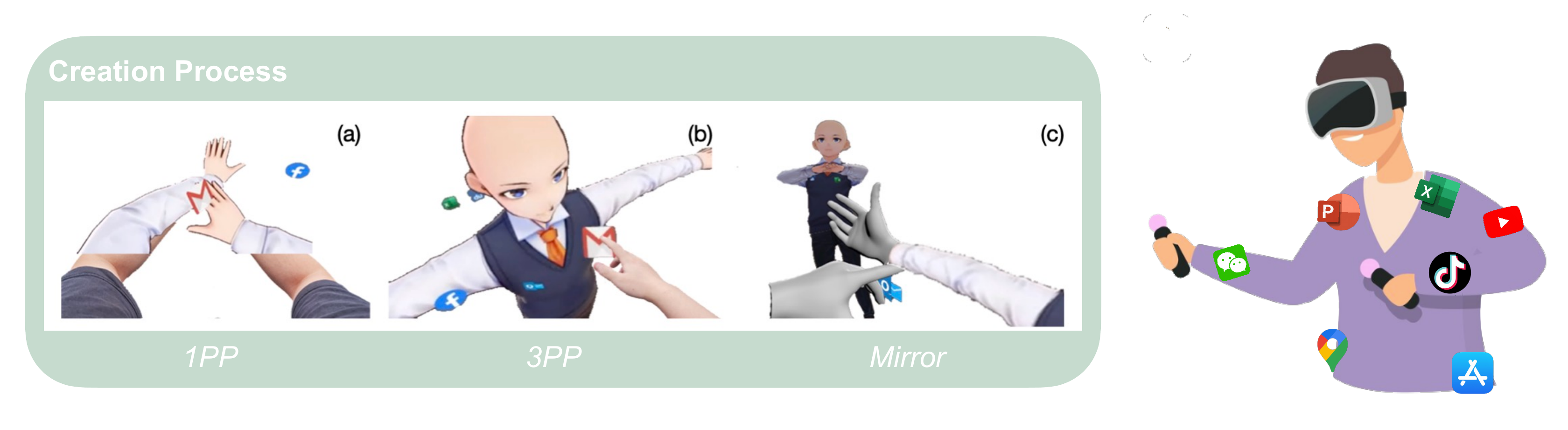}
  \caption{%
    Participants can customize their on-body menu creation process in VR using three perspectives: (a) first-person perspective (1PP), (b) third-person perspective (3PP), and (c) mirror perspective. 
  }
  \Description{This teaser figure shows participants customizing their on-body menu creation process in virtual reality (VR) using three different perspectives: (a) first-person perspective (1PP), where the user views their virtual arms directly in front of them; (b) third-person perspective (3PP) displaying an over-the-shoulder view of the user interacting with the menu on a virtual avatar; and (c) mirror perspective presenting the user’s avatar and menu in a mirror-like reflection. The right side of the image depicts a person using VR controllers and wearing a VR headset, surrounded by various app icons.}
  \label{fig:teaser}
\end{teaserfigure}


\maketitle

\section{Introduction}
Within the human-computer interaction (HCI) community, there is a strong interest in exploring how the human body drives the development of interactive technologies~\cite{harrison_-body_2012, harrison_implications_2014, harrison_skinput_2010,patibanda2023auto,patibanda2023fused,darbar2024onarmtext}. The user's body can provide a unified interface for users to directly access the information flow during interaction or entertainment, from interacting with digital media to ``experiencing our bodies as digital play''~\cite{mueller_experiencing_2018,floyd_mueller_limited_2021}. For example, Microsoft Xbox's Kinect\footnote{\href{https://en.wikipedia.org/wiki/Kinect}{https://en.wikipedia.org/wiki/Kinect}} as a body motion monitoring accessory makes body-based gaming possible, while wearable trackers support on-body gaming competitions.

These advantages of on-body interaction techniques present an alternative approach to designing interaction mechanisms in virtual reality (VR) and augmented reality (AR)~\cite{bergstrom_human--computer_2019,darbar2024onarmtext}, which involves using the user's body as a location (referred to as a bodily landmark~\cite{weigel_skinmarks_2017}) for placing menu items (referred to as on-body or body-referenced menus~\cite{lediaeva_evaluation_2020}). Users can leverage their embodied awareness to perceive and obtain information by intuitively interacting with their bodies in virtual environments. For instance, when wearing a VR headset, a user can extend their hand to select a menu item positioned on their forearm. This interaction is enabled by the user's innate awareness of their body's location in space and their ability to engage with the virtual environment using their body. Azai et al.~\cite{azai_tap-tap_2018}, for example, leveraged this to build a menu system appearing on various body parts, such as hands, arms, legs, and abdomen. Devising effective on-body menus, however, is a complex task, as it entails addressing numerous design challenges~\cite{lediaeva_evaluation_2020}, including the potential impact of different perspectives on user experience when creating on-body menus.

In this paper, we investigated the impact of the creation process on the practical application of on-body menus in VR. Prior studies have demonstrated that the perspective employed during information encoding influences recall accuracy~\cite{olszewska_encoding_2013}. There are still significant gaps in understanding how the design of on-body menus influences user performance and memory in VR environments. The impact of these design considerations on user experience and memory retention remains unclear. Additionally, the perception of avatars in VR plays a crucial role in shaping how individuals experience and remember virtual environments. Specifically, the avatar’s perspective can influence the sense of presence and spatial awareness, which may, in turn, affect memory accuracy and recall. Therefore, it is essential to explore both the creation of on-body menus and the effects of avatar perspective on representation and memory in VR to better inform the design of effective interfaces.

In this paper, we first provided three different approaches to users to build the on-body menus in VR: \textit{i.e.}, (1) the \textsc{First Person} Perspective \textsc{(1PP)}, (2) the \textsc{Third Person} Perspective \textsc{(3PP)}, and (3) the \textsc{Mirror} Perspective (see Figure \ref{fig:teaser}). Then, we conducted a within-subjects user study ($N$ = 12) comparing these three creation processes in terms of user experience (\textit{e.g.}, presence, enjoyment, and mental and physical load) and performance (\textit{e.g.}, precision and memorability). We compared these processes in VR in Creation and Recall tasks, where users were asked to first create a connection between 15 virtual icons and their bodies, and then recall the icons and their locations on the body after a short interval in the Recall task.

Our first study aimed to contribute to the understanding of the impact of the creation process on the practical application of on-body menus in VR and to provide guidance for designers and developers of VR applications. We found that the Mirror perspective had a faster creation time compared to 3PP, suggesting that a mixed perspective combining both viewpoints facilitated faster and more comprehensive menu creation, particularly when using a third-person perspective avatar. Both 3PP and Mirror perspectives exhibited faster recall times compared to 1PP, indicating that seeing the avatar or mirror reflection boosted users' confidence in recalling menu placement. 1PP resulted in a greater sense of involvement in the VR environment, as dragging icons directly onto the user's body enhanced the feeling of engagement and embodiment.

We also discovered that users tended to organize the placement of icons in on-body menus based on their category. To assess the generalizability of these initial findings, we conducted our second study ($N$ = 18) utilizing the latest VR equipment, Meta Quest 3, which offers an integrated body movement detection system without external sensors and higher accuracy with lighter weight. Following the same protocol as our first user study for the Creation Task, we required participants to place 15 icons from five categories using the three creation methods. We combined the collected distributions from both studies ($N$ = 30), which confirmed significant preferences in on-body menu distribution based on icon category. Specifically, social media icons were consistently placed on forearms, while productivity icons favored upper-body placement. We also found a significant association relationship between the Leisure and Social Media icon categories, as well as between Productivity and Social Media icons. 

Finally, we conducted comprehensive discussions on the bodily preferences associated with creating on-body menus in VR and synthesized design strategies. We believe these strategies can serve as valuable guidance for the development of effective on-body menus in VR environments, drawing from our research findings and insights. In summary, our contributions encompass:

\begin{itemize}

\item Development of an integrated on-body menu creation system devoid of external detection sensors or cameras, which can be easily replicated across the latest VR devices.

\item Investigation into the performance of various perspective-based creation processes for on-body menus in VR with two studies ($N$ = 12 and $N$ = 18).

\item Exploration of user preferences ($N$ = 30) regarding bodily landmarks within different application categories and analysis of the significance of their distribution.

\end{itemize}

\section{Related Work}
\label{related_work}
In this section, we examine research on on-body interactions in virtual environments. Additionally, we review the previous literature on graphical menu designs in virtual environments, including techniques for menu design and placement, as well as menu shape and format. Finally, we summarize the impact of perspective on memory and performance during interactions.

\subsection{Body-centric Interactions in VR}
On-body interaction techniques, which utilize the human body as an input/output platform, have been explored in various studies to enhance the immersion and naturalness of interactions in VR environments~\cite{harrison_-body_2012,bergstrom_human--computer_2019,coyle_i_2012} and provide more accurate eyes-free targeting~\cite{weigel_skinmarks_2017,gustafson_imaginary_2010}. Several studies have examined using different body parts, such as the arms, palms, and skin, as input surfaces for VR interactions~\cite{chatain_digiglo_2020,dezfuli_palmrc_2014,mistry_wuw_2009}. Skinput~\cite{harrison_skinput_2010} and Touché~\cite{sato_touche_2012}, for instance, explore using the body as an input surface by sensing acoustic and capacitive signals to recognize gestures on the skin. Meanwhile, Armstrong~\cite{li_armstrong_2021} examines the effectiveness of using non-dominant arm-anchored user interfaces (UIs) for pointing tasks in VR. 

Other studies have investigated the use of body-centric mid-air input surfaces~\cite{wagner_body-centric_2013}, such as Hand Range Interface~\cite{xu_hand_2018}, Swarm Manipulation~\cite{li2023swarm}, BodyLoci~\cite{fruchard_impact_2018} and BodyOn~\cite{yu_blending_2022}, to leverage bodily interfaces to achieve more effective mid-air interactions. Additionally, there have been works on exploring the use of on-body menus, such as Tap-tap~\cite{azai_tap-tap_2018} menu and PalmGesture~\cite{wang_palmgesture_2015}, and the role of visual and tactile cues in browsing them. These studies show the potential of on-body interaction in creating more intuitive and immersive VR experiences.

There is a gap, however, in understanding how these on-body interaction techniques influence the users' memory performance in VR. Therefore, one of our main motivations is to investigate how the design and implementation of on-body menus impact user memory performance and retention in VR environments.

\subsection{Graphical Menus in VR}

Graphical menus are a common user interface element in digital environments, including VR. In VR, graphical menus can be presented in various forms~\cite{bowman_design_2001}, such as floating menus~\cite{azai_tap-tap_2018}, hand-held menus~\cite{azai_open_2018}, or on-body menus~\cite{gerber_spin_2005}. The design of graphical menus in VR needs to consider several factors, such as user comfort, ease of use, and visibility~\cite{santos_comparative_2017}. Therefore, one common design strategy for graphical menus in VR is to place them in the user's field of view, usually at a distance and angle that minimizes visual strain and maximizes accessibility~\cite{bowman_introduction_2001}. For example, Gebhardt et al. place graphical menus at a fixed position relative to the user's headset, while others use dynamic menus that adapt to the user's gaze direction or hand movements~\cite{gebhardt_extended_2013}. Another design consideration for graphical menus is the layout and organization of menu items~\cite{lindlbauer_context-aware_2019}. Studies have shown that grouping menu items based on their functional category or relevance can improve user efficiency and satisfaction~\cite{dachselt_three-dimensional_2007}. Moreover, providing visual feedback when selecting menu items can help users perceive the selection process and reduce errors~\cite{xu_exploring_2020}.

However, the correlation between the specific design considerations of graphical menus and user performance remains underexplored. Our research is therefore motivated by the need to investigate how the design of graphical menus in VR impacts user performance and whether certain designs facilitate better performance.

\subsection{Perspectives and Memory}
The way in which an avatar is perceived within a virtual environment can have a significant impact on how people experience and remember that environment. Firstly, the avatar perspective can influence the sense of presence, or the feeling of ``being there,'' within the virtual environment~\cite{sanchez-vives_presence_2005} and levels of immersion~\cite{monteiro2018enjoyment} and engagement ~\cite{monteiro2018engagement}. Research has found that first-person perspectives generally lead to a higher sense of presence than third-person perspectives~\cite{denisova_first_2015,kallinen_presence_2007}. However, this can depend on whether participants are allowed to choose their avatars. In a study by Lim and Reeves~\cite{lim_being_2009}, participants who were allowed to choose their avatars reported a higher sense of presence in third-person perspectives than in first-person perspectives. This could be due to the increased sense of ownership that is experienced when the avatar becomes a proxy for the participant in the virtual environment. A stronger sense of presence during encoding can lead to greater memory accuracy~\cite{krokos_virtual_2019,makowski_being_2017}. Secondly, the avatar perspective can affect spatial awareness in virtual environments~\cite{gorisse_first-_2017,medeiros_keep_2018}. Third-person perspectives have been found to lead to faster response times and improved spatial awareness of information in the periphery of the scene~\cite{gorisse_first-_2017}. This could influence the types of details that people later recall, by increasing memory for peripheral information and/or the spatial layout of the overall scene~\cite{bonnail_memory_2023}.

Despite these insights, more research is needed to understand how the perspective of the avatar affects users' memory in VR. This gap in the literature motivates our research to explore the impact of the avatar perspective on memory performance and how it shapes users' recall of the virtual environment.

\section{On-body Menus in VR}

\subsection{Design On-body Menus in VR}

\begin{figure}[t]
  \centering
  \includegraphics[width=0.85\linewidth]{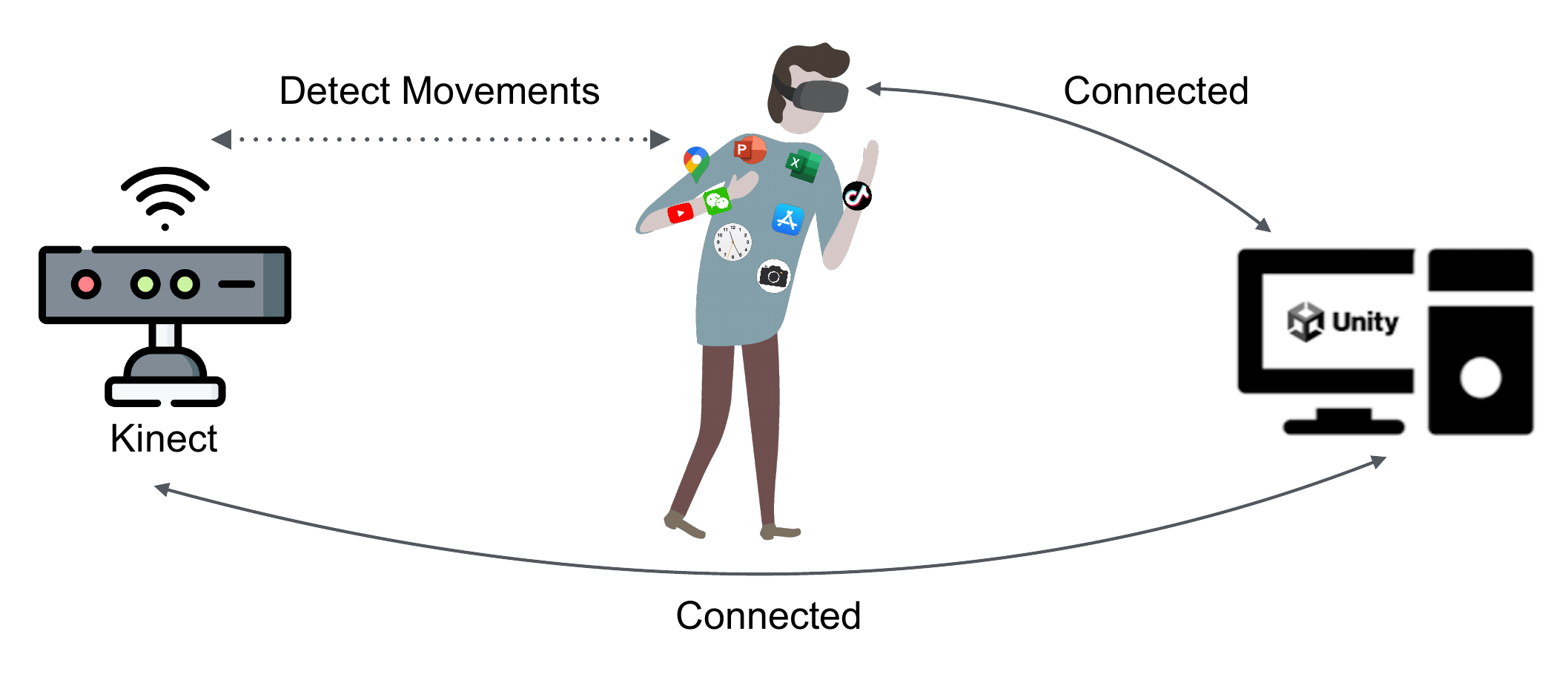}
  \caption{Our fist on-body menu system comprises three components: a VR headset (\textit{i.e.}, Oculus Rift S), a camera for whole-body motion detection (\textit{i.e.}, Kinect v2), and an interactive VR interface. The Kinect detects users' bodily movements, and the system exchanges signals among the devices to facilitate user interaction.}
  \Description{Diagram of a system showing the interaction between a Kinect device, a VR headset, and a computer running Unity. A person wearing a VR headset is depicted in the center, surrounded by icons representing various applications. The Kinect device is shown on the left with a dotted arrow labeled ``Detect Movement'' pointing towards the person. Two curved arrows labeled ``Connected'' indicate the bidirectional connections between the person and the computer running Unity on the right. The diagram illustrates how the Kinect detects the user's movements, which are then processed by the computer to facilitate interaction in the VR environment.}
  \label{fig:system}
\end{figure}

Similar to the Dock\footnote{\href{https://en.wikipedia.org/wiki/Dock_(macOS)}{https://en.wikipedia.org/wiki/Dock\_(macOS)}} provided by MacOS, we first envision an interaction scenario where a graphic functional interface is used in VR to quickly launch applications and to switch between running applications. Our on-body menu creation system comprises three components: a VR headset, a sensor for whole-body motion detection, such as cameras~\cite{li2021vrcaptcha} or IMUs~\cite{mollyn2023imuposer}, and a VR interface, and the system should exchange signals between the devices to facilitate user interaction (see Figure~\ref{fig:system}).

\subsection{Creation Processes}

As we mentioned in Section \ref{related_work}, the creation of differing perspectives in VR could have a profound impact on cognitive understanding. This involves constructing mental representations of information to facilitate later recall~\cite{peeters_misplacing_2019,lim_being_2009}. Particularly, this process of information linkage is crucial when creating on-body menus in VR. Users need to establish meaningful associations between virtual icons and their corresponding locations on the body. Therefore, when designing on-body menus in VR, it is vital to consider these factors in order to optimize the creation process and enhance the recall of memory. Hence, our motivation extends to exploring how the design and implementation of on-body menus, combined with different perspectives, can impact the users' cognitive understanding and memory recall in a VR environment. Thus, we provide three different creation processes to establish the on-body menus in VR: (1) \textsc{1PP}, (2) \textsc{3PP}, and (3) the \textsc{Mirror}.

\subsubsection{First-Person Perspective (1PP)}

The first-person perspective technique allows the user to see a first-person avatar in the VR environment and interact with virtual icons placed directly in front of them. Users can manipulate these icons directly, for example, by dragging an icon to a specific body landmark to establish a connection (see Figure \ref{fig:teaser} (a)). This technique takes advantage of the user's proprioceptive sense, which is the ability to perceive the position and movement of one's body parts without relying mainly on visual feedback. This technique could also provide a more immersive experience as the user can directly see and manipulate the virtual icons with their hands, similar to real-world interactions. However, the disadvantage of this technique is that the user's arms and hands may obstruct their view of the virtual environment, which can decrease spatial awareness and increase mental workload. Additionally, the user may experience discomfort or fatigue due to prolonged arm movements during the creation process.

\subsubsection{Third-Person Perspective (3PP)}

The third-person perspective technique allows users to see their T-pose avatars in the VR environment with virtual icons attached to their bodies. Users can use their controllers or hands to establish connections between the virtual icons and landmarks on the 3PP avatar, similar to the 1PP technique (see Figure \ref{fig:teaser} (b)). The advantage of this technique is that users can have a better overall view of the virtual environment and spatial awareness since their arms and hands do not obstruct their view. However, this technique may require more mental workload and lead to split attention effects \cite{sweller2011split}, as users must establish connections between the virtual icons and the landmarks on the avatar while relying on the visual feedback of the avatar and the spatial relationship of the virtual icons. Moreover, users may not feel as immersed in the virtual environment as in the 1PP technique due to the lack of direct hands-on interaction with virtual icons.

\subsubsection{Mirror Perspective}

In the Mirror perspective technique, users see a first-person avatar in the VR environment and can manipulate virtual icons directly in front of them, as in the 1PP technique. However, a virtual mirror is also present in the environment, which allows users to see their entire body and the location of the virtual icons on it from a third-person perspective (see Figure \ref{fig:teaser} (c)). This technique takes advantage of both the proprioceptive sense and the visual feedback provided by the mirror to improve spatial awareness and understanding of the on-body menu layout~\cite{li_vmirror_2021}. The mirror can also help to reduce the user's cognitive load by providing an external representation of their body and the virtual icons attached to it. However, the use of the mirror may also increase the user's mental workload, as they have to constantly switch between first-person and third-person perspectives to establish connections between virtual icons and body landmarks. Additionally, the presence of the mirror perspective may cause visual distractions, reducing the sense of immersion in VR.

\section{User Study 1: Evaluate Three Creation Processes of Creating On-body Menus in VR}
\label{User Study}

First, we carried out a user study to explore the three on-body creation processes in VR in terms of performance (\textit{i.e.}, precision, and memorability) and user experience. The study employed a within-subjects experimental design with one independent variable, the \textsc{Creation Process}, consisting of three techniques: \textsc{1PP}, \textsc{3PP}, and \textsc{Mirror}. To mitigate any learning effects, the Creation Process order for participants was counterbalanced using a Latin square approach. The experiment spanned 3 days per condition, with each participant dedicating 20 to 30 minutes daily to both Creation and Recall Tasks, consistent with Bergstrom et al.~\cite{bergstrom-lehtovirta_placing_2017}. The objective of User Study 1 is to address \textbf{\textit{RQ1: How do different perspective-based creation processes impact user experience and usability?}}

\subsection{Procedure}
Before the experiment, participants were asked to complete the demographic form. After each test, participants completed a questionnaire to collect feedback about their experience and other subjective data. Besides, we added the final comparison and asked participants to rate their preference on a 7-point Likert scale and summarize their strategies and suggestions via a semi-structured interview.

In the Creation Task, participants first received a tutorial and demo video to familiarize themselves with the on-body mapping techniques assigned to them. Afterwards, they wore the VR headset and underwent a 15-minute tutorial and training session for the creation process. Once participants confirmed their familiarity with the techniques, they proceeded to the formal Creation Task. During this process, participants were instructed to create an on-body menu using the assigned creation process. They were asked to connect 15 virtual icons to their body landmarks, and the order of icon presentation was random. To better understand the post-creation distribution, we categorized the icons into five groups, and all participants were familiar with those applications' icons (see Table~\ref{tab:icon_categories}). After creating their on-body menus, they were given 1 minute to preview the distribution of icons on their bodies, \textit{i.e.}, the post-creation on-body menu. Participants then filled out questionnaires and took a 5-minute break before moving on to the Recall Task.

\begin{table}[ht]
  \centering
  \caption{Categories of Icons. The images depicting these icons can be found in Figure \ref{fig:icons}.}
  \label{tab:icon_categories}
  \begin{tabular}{cc}
    \toprule
    \textbf{Category} & \textbf{Icons} \\
    \midrule
    \textit{Social Media} & Weibo, WeChat, QQ \\
    \midrule
    \textit{Productivity} & Excel, Word, PowerPoint \\
    \midrule
    Leisure & BiliBili, YouTube, TikTok \\
    \midrule
    \textit{Utilities} & Camera, AppStore, Clock \\
    \midrule
    \textit{Other} & Google Maps, Ele.me, Uber \\
    \bottomrule
    \Description{Table titled ``Categories of Icons'' and the table contains two columns: Category and Icons. The categories are Social Media, Productivity, Leisure, Utilities, and Other. Under Social Media, the icons listed are Weibo, WeChat, and QQ. Under Productivity, the icons listed are Excel, Word, and PowerPoint. Under Leisure, the icons listed are BiliBili, YouTube, and TikTok. Under Utilities, the icons listed are Camera, AppStore, and Clock. Under Other, the icons listed are Google Maps, Ele.me, and Uber.}
  \end{tabular}
\end{table}

In the Recall Task, participants were also given a tutorial and demo video to read and watch before starting the recall test. A panel appeared in front of them and provided icons randomly, and participants were asked to recall the corresponding position they had created earlier and tap the corresponding body landmark as accurately as possible and within a reasonable time.

\subsection{Participants and Apparatus}

A total of 12 participants (8 females and 4 males) were recruited from a local university. The age range of the participants was between 19 and 30 years ($M = 21.5, SD = 2.88$). Ten participants are students, and two are research scientists or faculty members of this university. All participants reported previous experience with VR, with familiarity ratings ranging from 2 to 7 on a 7-point Likert scale, where 1 means no experience in VR and 7 means an expert ($M = 4.42, SD = 1.51$).

This system was built with Unity 2019.3.18f, and we used the Kinect to detect users' bodily movements. The Oculus Rift S VR headset was chosen for the study due to its capability to provide real-time detection from the Kinect and its compatibility with our system. Data collection was done using a PC with Windows 10 operating system, which was also connected to the Rift S.

\subsection{Measures}

We used several measures to evaluate the effectiveness and usability of the three creation processes in the Creation Task. Our dependent variables included \textit{Presence}, which was measured with the Igroup Presence Questionnaire (IPQ)~\cite{schubert_experience_2001}; \textit{Enjoyment}, which was measured with a single 7-point Likert scale; \textit{Perceived Mental load and Physical Demands}, which were measured with the unranked NASA-TLX~\cite{hart_nasa-task_2006}; and \textit{Confidence in Memorability}, which was measured with a single 7-point Likert scale. We also collected their \textit{Time Cost} as the time taken to create the on-body menu and \textit{Post-creation Distribution}, which was automatically saved after the users confirmed at the end of the creation process.

We also used several measures to evaluate users' memorability of the on-body menus created in the Recall Task. Dependent variables included \textit{Error Distance} (or Error Magnitude~\cite{zhai_silk_1994}), measured as Euclidean summation, where we calculate the distance from the centre of the circle where the user placed the icon in the Creation Task to the position where the user tried to tap in the Recall Task, and \textit{Recall Time}. Participants were also asked to complete the unranked NASA-TLX~\cite{hart_nasa-task_2006} questionnaire to measure \textit{Mental and Physical Demands} and \textit{Performance}.

\subsection{Results: Creation Task}
\label{results}

\begin{figure*}[ht]
  \centering
  \includegraphics[width=\linewidth]{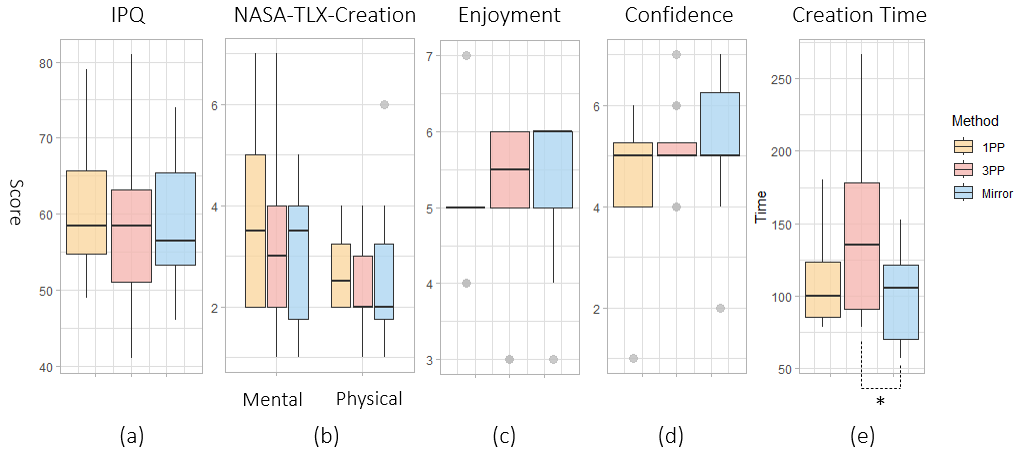}
  \caption{Results of IPQ (a), NASA-TLX (Mental and Physical Demands) (b), Enjoyment (c), Confidence in Memorability (d), and Creation Time (e) for the three techniques during the Creation Task, with `*' indicating $p < 0.05$.}
  \Description{Five boxplots comparing the results of three techniques (1PP, 3PP, and Mirror) for various measures during the Creation Task. The measures are IPQ, NASA-TLX (Mental and Physical Demands), Enjoyment, Confidence in Memorability, and Creation Time. Each measure is represented by a separate boxplot labeled (a) through (e).}
  \label{fig:creation}
\end{figure*}

\subsubsection{Presence}
A one-way repeated measures ANOVA was utilized to assess the impact of \textit{Techniques} (1PP, 3PP, and Mirror) on presence scores, as determined by the IPQ. The analysis did not yield a significant effect of the techniques on presence scores ($F(2, 22) = 1.38$, $p = 0.27$).

Beyond the overall Presence scores, the INV (Involvement) subscale of the IPQ was analyzed to assess cognitive and emotional engagement among participants. The analysis showed a significant effect within this subscale, with the INV score for 1PP ($M = 17.42$, $SD = 3.73$) being significantly higher compared to 3PP ($M = 15.33$, $SD = 3.34$, $p < 0.05$). No significant difference was found between the INV scores for 1PP and Mirror ($M = 17.33$, $SD = 3.17$, $p = 0.89$).

\subsubsection{Mental and Physical Demands}
The ANOVA indicated no statistically significant differences in self-reported mental demands across the techniques ($F(2, 22) = 0.57$, $p = 0.57$, generalized $\eta^2 = 0.02$). Similarly, no significant differences were found in the Physical Demands subscale ($F(2, 22) = 0.85$, $p = 0.44$, generalized $\eta^2 = 0.03$).

\subsubsection{Enjoyment}
A statistical analysis was performed to evaluate the participants’ reported levels of enjoyment following the completion of the sessions involving all three techniques. The results from a one-way ANOVA revealed no statistically significant differences in enjoyment across the techniques ($F(2, 22) = 0.73$, $p = 0.49$, generalized $\eta^2 = 0.02$).

\subsubsection{Confidence in Memorability}
The analysis did not reveal a statistically significant difference in participants’ confidence regarding the memorability associated with the three techniques ($F(2, 22) = 1.42$, $p = 0.26$, generalized $\eta^2 = 0.17$).

\subsubsection{Time Cost}
The analysis of variance revealed a statistically significant difference in time cost among the techniques ($F(2, 22) = 4.03$, $p < 0.05$, generalized $\eta^2 = 0.17$). Subsequent pairwise comparisons, adjusted by Bonferroni correction, indicated that the time cost for 3PP ($M = 147.57$, $SD = 66.73$) was significantly higher than for Mirror ($M = 101.83$, $SD = 32.82$, $p < 0.05$). No significant difference in time cost was observed between 1PP and Mirror techniques ($p = 0.54$) or between 1PP and 3PP ($p = 0.07$).

\subsection{Results: Recall Task}

\begin{figure}[t]
  \centering
  \includegraphics[width=0.75\linewidth]{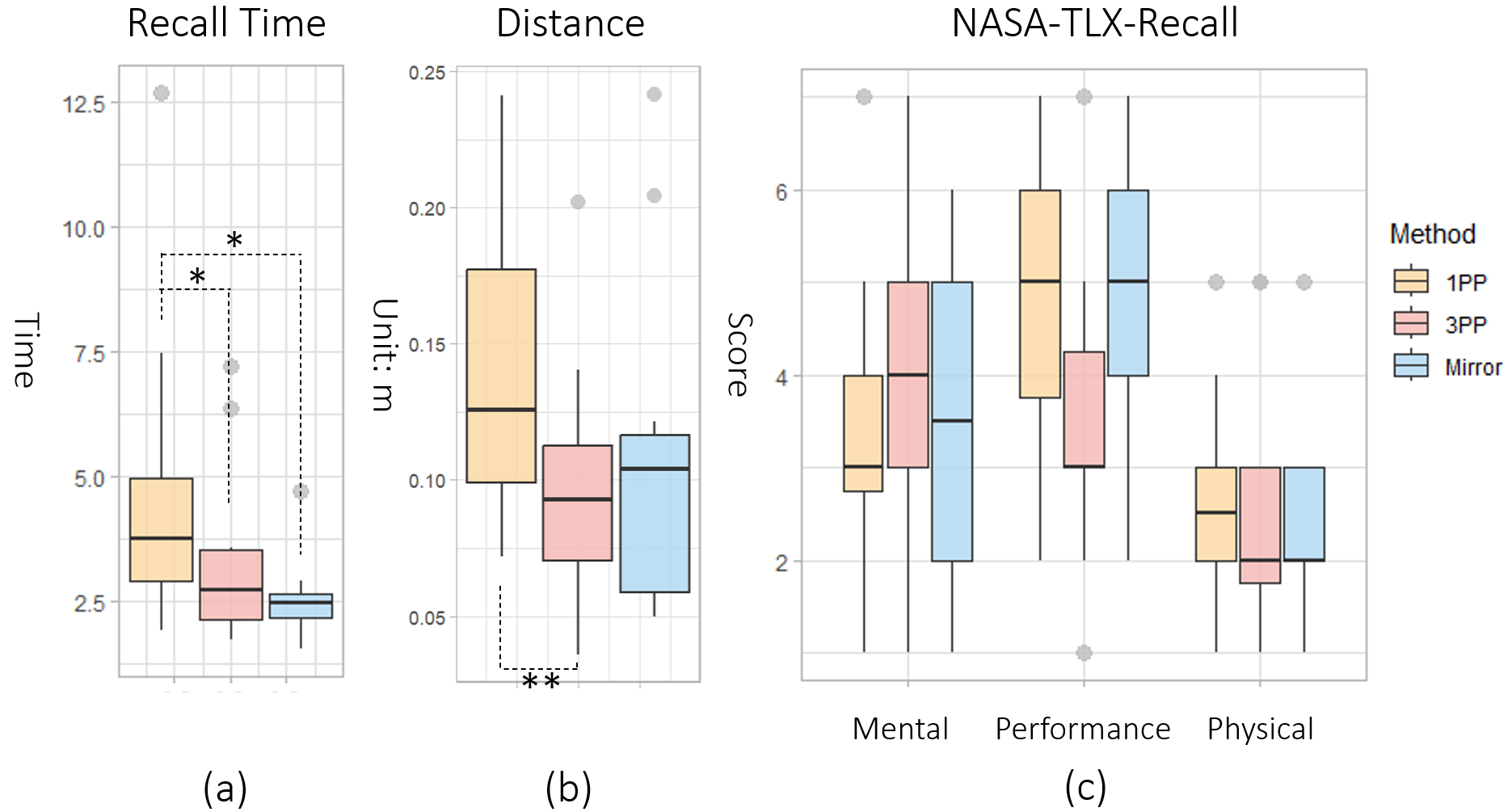}
  \caption{Results of Recall Time (a), Error Distance (b), and NASA-TLX (Mental and Physical Demands and Performance) (c) for three techniques during the Recall task, with `*' and `**' indicating $p < 0.05$ and $p < 0.01$, accordingly.}
  \Description{Three boxplots comparing the results of three techniques (1PP, 3PP, and Mirror) for various measures during the Recall task. The measures are Recall Time, Error Distance, and NASA-TLX (Mental and Physical Demands and Performance). Each measure is represented by a separate boxplot labeled (a) through (c).}
  \label{fig:recall}
\end{figure}

\subsubsection{Recall Time}
A one-way repeated measures ANOVA was conducted to assess the impact of three distinct \textit{Techniques} on participants’ \textit{Recall Time}. Mauchly’s test confirmed the assumption of sphericity ($\chi^2(2) = 0.66$, $p = 0.13$), so no corrections were necessary. The results indicated a significant main effect of Techniques on recall time ($F(2, 22) = 7.85$, $p < 0.01$). Post-hoc analyses with Bonferroni correction showed that recall times were significantly longer for 1PP ($M = 4.64$, $SD = 2.95$) compared to 3PP ($M = 3.24$, $SD = 1.77$, $p < 0.05$) and Mirror ($M = 2.51$, $SD = 0.80$, $p < 0.05$). No significant difference was found between 3PP and Mirror ($p = 0.09$).

\subsubsection{Mental and Physical Demands \& Performance}
A one-way repeated measures ANOVA was employed for analysis. The Mental Demand subscale scores did not show significant differences across techniques ($F(2, 22) = 1.32$, $p = 0.287$). In the Physical Demand subscale, no significant differences were detected across the techniques ($F(2, 22) = 0.585$, $p = 0.501$) following the Greenhouse-Geisser correction. Regarding the Performance subscale, the ANOVA yielded non-significant results ($F(2, 22) = 2.970$, $p = 0.07$).

\subsubsection{Final Preference}
In terms of final preference, most participants ($N$ = 9) ranked Mirror as the best technique for memorizing icon distribution. They found that Mirror provided a direct and intuitive way to visualize the relative positions of icons on the body. This creation process allowed users to see the exact location of icons on their body and make adjustments in real-time, which facilitated memorization. P4 noted, \textit{``Mirror allows for a clearer view of the icons' relative positions on the body. In 1PP, it is challenging to see the entire icon layout on the body due to the viewing angle, and 3PP lacks the simplicity and intuitiveness of looking directly in the mirror.''} 

Some participants ($N$ = 4) found 3PP beneficial as it offered a better viewing angle and enabled the manipulation and rotation of the avatar model, aiding in memory formation. However, P10 noted that 3PP was more complicated and challenging to use compared to the other two techniques, as it was hard to map the positions of the icons on the avatar to their own body. \textit{``3PP allows users to see all placement positions directly, making it the best. However, it is difficult to find the correspondence between 3PP and one's own body.''}  

Meanwhile, 1PP was often ranked as the worst technique by participants ($N$ = 7). The main challenge with 1PP was the difficulty in accurately placing icons on the body due to the limited field of view. In VR, users could not directly see the placement of icons, leading to potential inaccuracies. While some participants mentioned that the physical feedback from the body could assist with memorization, the lack of visual references hindered the process. P1 suggested, \textit{``Consider taking into account the height and body shape of the test participants.''} This highlights the importance of avatar diversity in embodiment research, as varying avatars’ body shapes and sizes could influence how effectively users interact with and remember icon placements~\cite{spiel2021bodies}.
\begin{figure}[t]
  \centering
  \includegraphics[width=\linewidth]{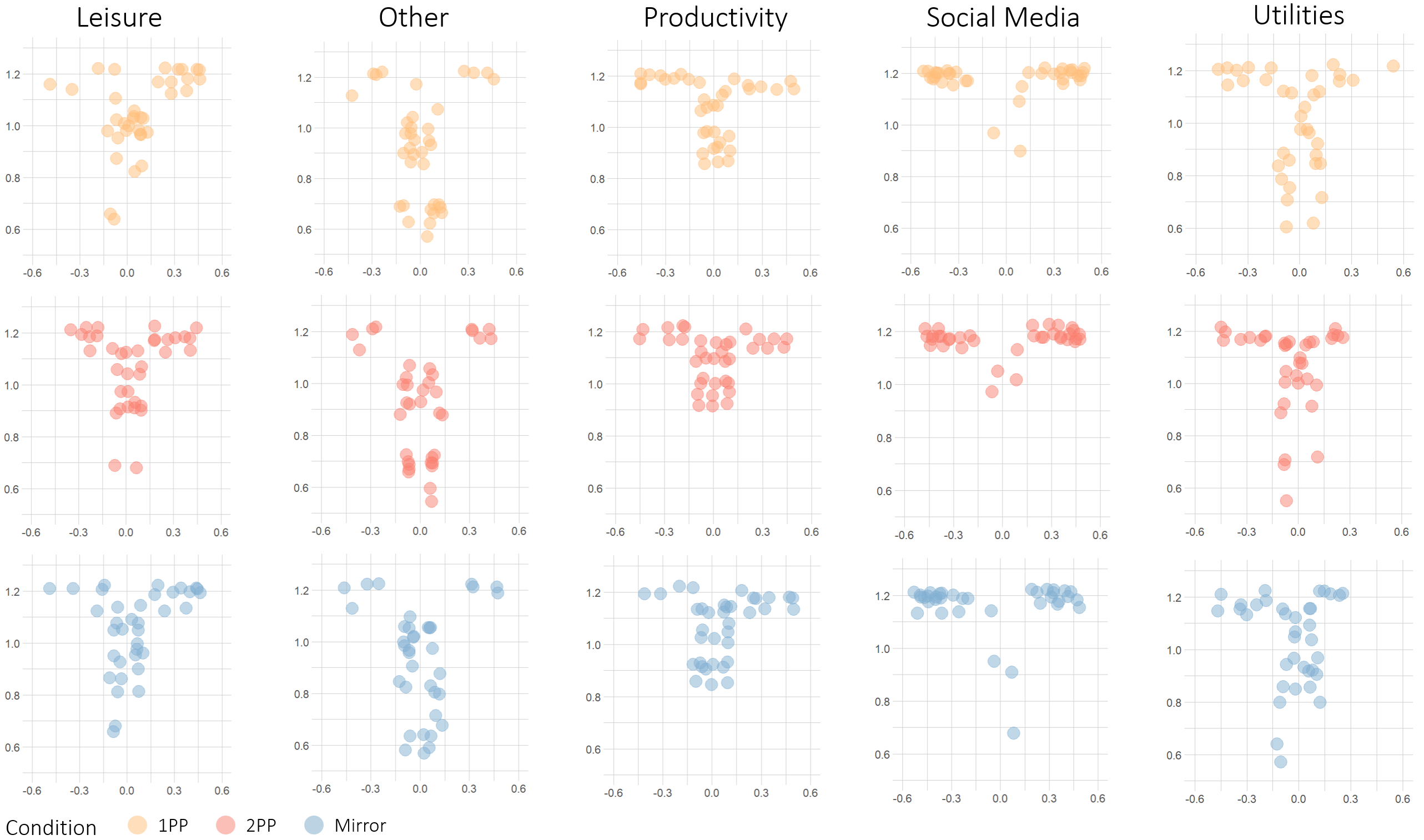}
  \caption{The distribution of different categories of icons under different conditions in User Study 1. The x and y coordinates in this figure represent unit lengths in Unity.}
  \Description{A series of scatter plots showing the distribution of different categories of icons (Leisure, Other, Productivity, Social Media, Utilities) under three conditions (1PP, 3PP, Mirror) in User Study 1. Each column represents a category of icons, and each row represents a condition. The first row contains scatter plots for 1PP (yellow), showing the distribution of icons in Leisure, Other, Productivity, Social Media, and Utilities. The second row contains scatter plots for 3PP (pink), showing the distribution of icons in the same categories. The third row contains scatter plots for Mirror (blue), also showing the distribution of icons in the same categories. Each scatter plot depicts data points indicating where the icons fall within the category for each condition, highlighting any clustering or spread in the distribution.}
  \label{fig:dis-category}
\end{figure}

\subsection{Icon Distribution}

We visualized the distribution of icons across various categories under different conditions (see Figure~\ref{fig:dis-category}). Interestingly, we observed a potential correlation between the location of icons on the body (referred to as landmarks) and their respective categories. For example, in Figure~\ref{fig:dis-category}, Productivity icons consistently avoid placement on the lower body, while social media icons tend to be predominantly positioned on the hands, with only scattered instances on the torso. Similar patterns were noted for icons in other categories, suggesting potential user preferences and distribution rules.

\section{User Study 2: Explore Bodily Preferences and Icon Distributions}

Our initial observations on icon distribution led us to conduct a second user study to expand our dataset and investigate the relationship between icon categories and body placement more rigorously. We employed a one-way within-subjects design similar to our first study. The independent variable, \textsc{Creation Process}, remained at three levels: 1PP, 3PP, and \textsc{Mirror}. Counterbalancing of the techniques was used to mitigate order effects. Participants completed all three creation processes with our latest system in a single day, with a 5-minute break provided between each condition. The entire study took under 40 minutes. The purpose of User Study 2 is to address \textbf{\textit{RQ2: Do users exhibit consistent patterns in the placement of icons across different body landmarks and categories?}}

\subsection{Procedure}

Before starting, participants provided demographic information and were given a tutorial and demonstration video, followed by a 10-minute training session. During the study, they created on-body menus using the same 15 virtual icons as in our initial study. Participants were asked to position the icons thoughtfully and could refine their placements until they were confident in their memorization. Once participants confirmed their placement, they would receive visual feedback and could then start placing the next icon or adjusting the previous ones. When using a different perspective-based creation process, users were told that they were free to adjust the distribution of the 15 icons on their body. After each task, they discussed their placement strategy and completed a Post-Creation Questionnaire.

\subsection{Participants and Apparatus}

We recruited 18 university students (12 males, 5 females, and 1 who preferred not to say) from a local institution. The participants' ages ranged between 19 and 36 years ($M = 23.72, SD = 4.18$). All students indicated previous experience using VR, with familiarity ratings between 2 and 7 ($M = 3.44, SD = 1.46$). None of them participated in our first study.

In our first study, participants noted that the Rift S headset's display and weight and the Kinect's unreliable body movement detection were obstacles for users. To address these concerns, we developed a new creation system utilizing the Meta Quest 3 headset, which provides the capacity of integrated body movement detection and wireless connection. On the other hand, using the built-in cameras on the Quest 3 as the input for motion capture enhances the replicability of the research. However, as a trade-off, the current Quest 3’s detection range does not offer full-body tracking like a standalone Kinect placed in front of the user. Instead, it requires users to move their heads accordingly while wearing the headset to allow the cameras to capture hand and body movements effectively. The new system was built with Unity 2021.3.9f1.

\subsection{Measures}

Consistently, we measured similar user experience metrics as in the first study, which included \textit{Presence}, \textit{Enjoyment}, and \textit{Perceived Mental and Physical Demands}. Additionally, we used the System Usability Scale (SUS)~\cite{brooke1996sus} to evaluate the feasibility and usability of our latest integrated creation system for on-body menus in VR. 

\subsection{Results}

\subsubsection{Presence}
A one-way repeated measures ANOVA was conducted to evaluate the effect of different creation \textit{Techniques} on presence scores, as measured by the IPQ. The average presence scores were $M = 59.44$ ($SD = 7.12$) for 1PP, $M = 60.22$ ($SD = 8.38$) for 3PP, and $M = 59.11$ ($SD = 9.79$) for Mirror. The analysis did not yield a significant effect of the techniques on overall presence scores ($F(2, 22) = 0.081, p = 0.920$).

\subsubsection{Mental and Physical Demands}
The perceived mental demands were $M = 2.67$ ($SD = 1.68$) for Mirror, $M = 2.28$ ($SD = 1.23$) for 1PP, and $M = 2.11$ ($SD = 1.28$) for 3PP. The physical demands were $M = 2.56$ ($SD = 1.58$) for 1PP, $M = 2.39$ ($SD = 1.61$) for Mirror, and $M = 1.89$ ($SD = 0.90$) for 3PP. However, the ANOVA analysis indicated that there were no significant differences between techniques on both Mental Demands ($F(2, 22) = 0.736, p = 0.484$) and Physical Demands ($F(2, 22) = 1.100, p = 0.341$).

\subsubsection{Enjoyment}
The mean enjoyment scores were $M = 5.78$ ($SD = 1.06$) for 3PP, and $M = 5.39$ ($SD = 1.24$) for Mirror and $M = 5.39$ ($SD = 1.20$) for 1PP. However, the results from a one-way ANOVA indicated no statistically significant differences in enjoyment between the techniques ($F(2, 28) = 0.664, p = 0.519$).

\subsubsection{System Usability Scale}
The 3PP condition had a perceived usability score of $M = 75.83$ ($SD = 11.76$), the Mirror condition had $M = 72.92$ ($SD = 14.04$), and the 1PP condition had $M = 65.83$ ($SD = 12.57$). The ANOVA analysis did not reveal a statistically significant difference in perceived usability among the conditions ($F(2, 22) = 2.89, p = 0.065$). While both 3PP and Mirror conditions scored above the average threshold of 68, suggesting generally acceptable usability, it is noteworthy that the 1PP condition scored below this threshold, indicating potential usability challenges~\cite{brooke1996sus}.

\subsection{Towards Understanding Icon Distributions}

\begin{figure}[t]
    \centering
    \includegraphics[width=0.75\linewidth]{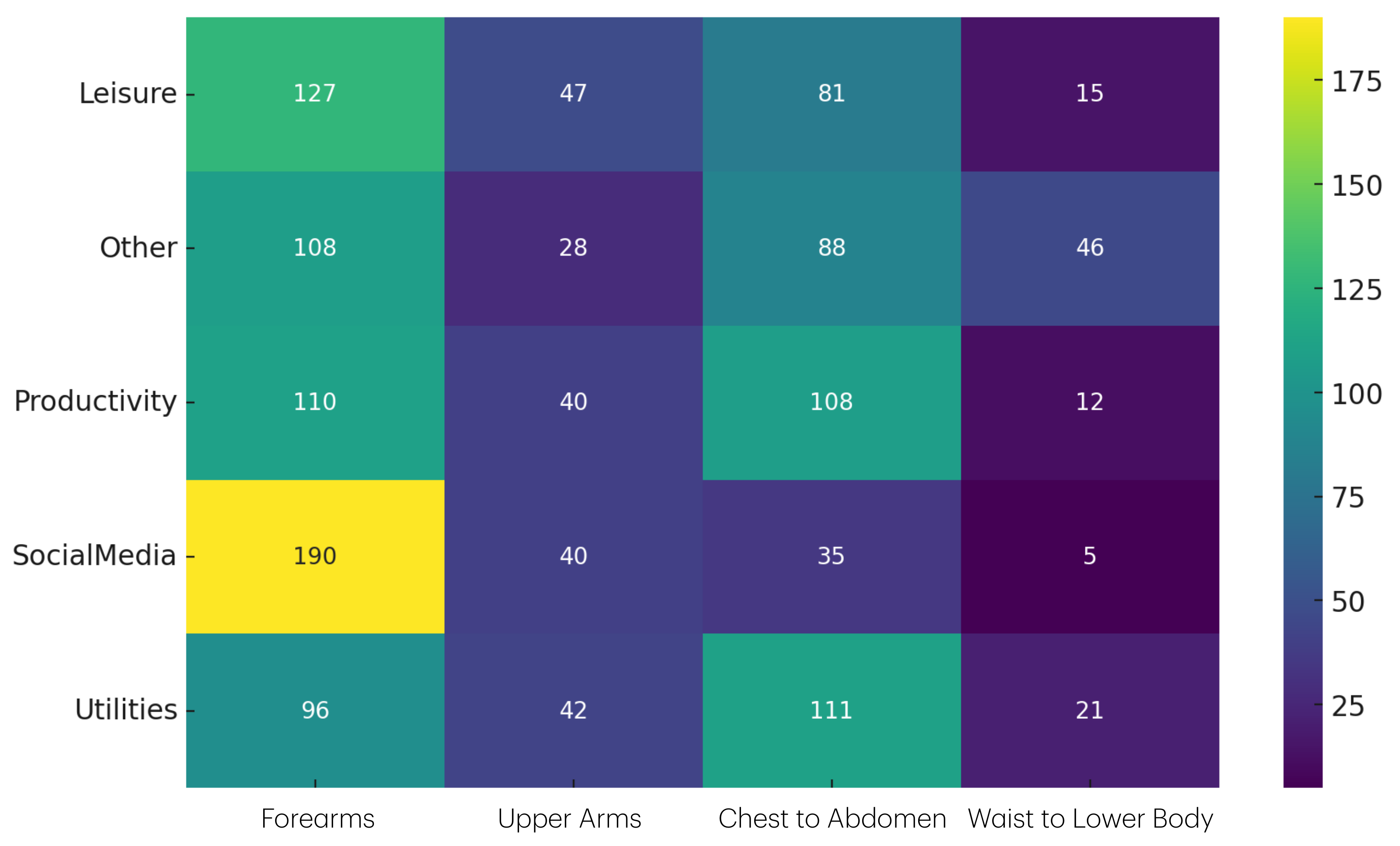}
    \caption{Heatmaps of icon distribution across areas, where each cell's intensity represents the count of icons, representing the relative frequency of categories across different areas.}
    \Description{A heatmap showing the distribution of icon categories (Leisure, Other, Productivity, Social Media, Utilities) across different areas (1 to 4). Each cell's intensity represents the count of icons, with higher counts shown in brighter colors and lower counts in darker colors. The categories are listed on the y-axis, and the areas are listed on the x-axis. The heatmap visually represents the relative frequency of each icon category across different areas, with a color scale bar indicating the range from lower to higher counts.}
    \label{fig:heatmap}
\end{figure}

\begin{figure}[t]
    \centering
    \includegraphics[width=\textwidth]{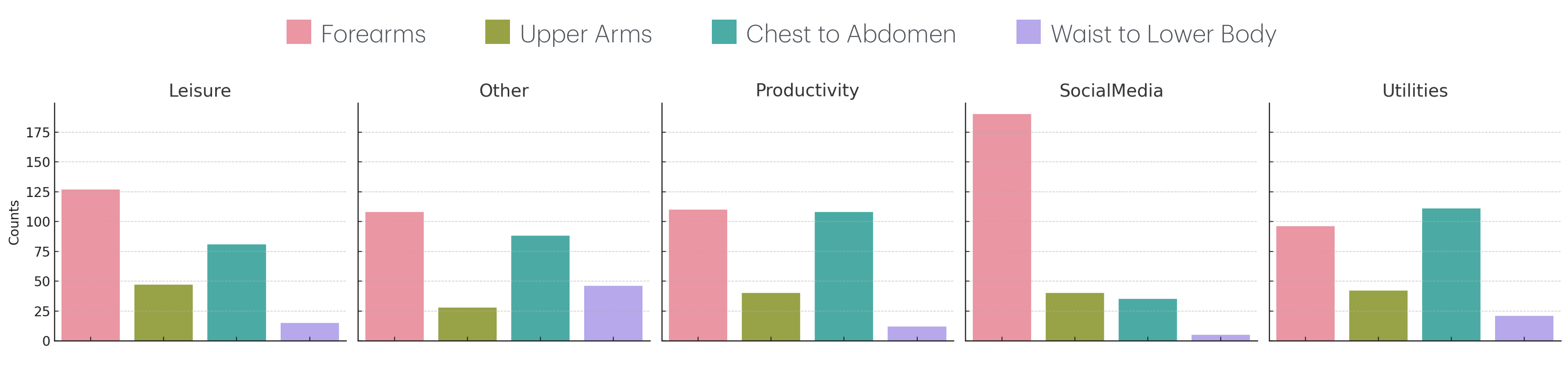}
    \caption{The distribution of each icon category across different body areas in User Study 2.}
    \Description{A series of bar charts showing the distribution of each icon category (Leisure, Other, Productivity, Social Media, Utilities) across different body areas in User Study 2. Each chart represents a category of icons, with the x-axis labeled ``Area'' (ranging from 1 to 4) and the y-axis labeled ``Counts''. Each bar chart visually depicts the number of icons in each category placed in different body areas, highlighting any significant differences in distribution.}
    \label{fig:bar_plots}
\end{figure}

To better understand the preference for bodily landmarks, we categorized the body into specific regions: the forearm, upper arm, chest to abdomen, and waist to lower body, inspired by \cite{fruchard_impact_2018}. We merged the icon distributions from our first and second studies. We used Z-tests to measure the proportions and the Apriori algorithm for association rule mining \cite{han2000mining,agrawal1994fast}, alongside confidence and lift metrics, to examine the association relationships among these categories. Bar plots illustrating the distribution of icons across areas (see Figure \ref{fig:bar_plots}) and heatmaps (see Figure \ref{fig:heatmap}) were generated for visual assessment of the distribution.

Our analysis revealed significant preferences in icon placement based on their functional categories. Social Media icons were overrepresented on the forearm with $Z = 5.54$ ($p < 0.0001$). Conversely, these icons were less likely to be positioned on the chest to abdomen and waist to lower body, with both $p < 0.0001$. Icons in the Other category showed a significant preference for the waist to lower body region, with $Z = 5.28$ ($p < 0.0001$). Utility icons were more likely to be placed on the chest to the abdomen with $Z = 2.92$ ($p = 0.0035$) and less likely to be situated on the forearm with $Z = -3.35$ ($p = 0.0008$). Similarly, Productivity icons were more likely to appear in the chest to abdomen region with $Z = 2.61$ ($p = 0.009$) and less likely in the waist to lower body area with $Z = -2.40$ ($p = 0.0163$).

We discovered significant relationships in our analysis of the associations between different icon categories across various conditions. Specifically, we found an association between the Leisure and Social Media icon categories, as well as between Productivity and Social Media icons. The association rule between Leisure and Social Media exhibited low support ($Support = 32.5\%$) but high confidence in the direction of Leisure predicting Social Media ($Confidence = 65.0\%$). The lift value ($lift = 1.23$) indicates that these two categories co-occur 23\% more frequently than would be expected by chance alone. This suggests a strong connection in user behavior. Similarly, the rule linking Productivity and Social Media demonstrated lower support ($Support = 25.4\%$) with moderate confidence values in both directions: Productivity predicting Social Media ($Confidence = 50.71\%$) and Social Media predicting Productivity ($Confidence = 58.68\%$). The lift value ($lift = 1.17$) further suggests that these categories co-occur more frequently than would be expected by chance.

\section{Discussion, Limitations, and Future Work}

Firstly, it must be acknowledged that our findings are not \textit{revolutionary} but rather \textit{incremental}. Our first user study addresses \textbf{RQ1}, revealing that the three creation processes have varying impacts on user experience and usability. While most participants preferred the Mirror condition due to its speed in creation and recall, their overall preferences among the three conditions were not consistent. Our second user study investigates \textbf{RQ2}, showing that participants tend to favor placing specific categories of icons in particular locations.

Therefore, our research on developing on-body menus in VR has unveiled two principal design strategies that can enhance user experience and performance in future implementations~\cite{mueller_towards_2023}. First, our analysis of on-body menu creation revealed insightful patterns in participants' preferences for organizing application icons across different body regions. Specifically, we observed a trend where Social Media icons are predominantly placed on the forearm, likely due to their familiarity with and frequent use of these applications for quick access. Conversely, icons related to Utilities and Productivity are generally positioned from the chest to the abdomen, suggesting a preference for these icons to be in a more visible location during task-focused interactions. Given this, designers could provide suggestions to users when setting up on-body icons. Secondly, it is advantageous for designers to incorporate customizable options for on-body menu configurations. Although our study indicated that the three perspective-based creation processes did not always significantly differ in quantitative performance outcomes, individual preferences for these setups were evident. Offering customization options, such as perspective toggles, can cater more effectively to user-specific needs in VR on-body menu creation.

Our study primarily investigated the associations between icon categories and their placement on the body, acknowledging other crucial factors that influence the efficacy of VR graphical menus. These factors include the size and visual complexity of the menu items, which can affect readability and the ease of selection, as well as the various input modalities (\textit{e.g.}, hand-tracking, controllers, gaze-based controls). Each input type may require specific design modifications to optimize usability and comfort. Nevertheless, we contend that the observed correlations and associations regarding icon placement are valuable independently and should remain pertinent even as input methods or icon characteristics evolve.

Furthermore, our research was limited to a fixed icon size and quantity. Future studies should explore the optimal icon sizes and numbers that users find comfortable and practical for on-body VR menus. This could include examining how changes in icon scale or density within the menu affect user experience. Additionally, investigating individual differences in spatial reasoning and visual preference could enhance our understanding of effective on-body menu design. These future inquiries will provide essential insights that complement our current findings on icon category distribution. Finally, we acknowledge that the sample size in both studies is not significant, and the impact of frequency of use and familiarity with using these applications (icons) could have a potential effect on the results.

Future work could focus on designing more integrated systems for whole-body menus in VR, such as leveraging only controllers and headsets \cite{ahuja2022controllerpose} or using other simple sparse inertial sensors \cite{9879760,winkler22} within a reinforcement learning framework to predict full-body movements. This approach would enable a more accurate investigation of icon distribution across the entire body and allow testing with a broader range of icon categories.

\section{Conclusion}

Our exploration into the impact of different creation processes on on-body menus in virtual reality (VR) has yielded several significant contributions and insightful discoveries. In our first user study ($N$ = 12), the use of the Mirror perspective was found to facilitate quicker creation times and improve recall accuracy compared to First Person Perspective (1PP) and Third Person Perspective (3PP). These findings support a mixed-perspective approach that combines the immediacy of a first-person view with the comprehensive visibility provided by a mirror or avatar. This strategy has the potential to significantly enhance both the efficiency and memorability of on-body menu design.

In our second study ($N$ = 18), we enhanced our system with an integrated body movement detection system using Quest 3 and recruited additional participants to expand our dataset on icon distribution for on-body menus. Our research consistently demonstrated patterns in user preferences for icon placement across different body landmarks. For instance, Social Media icons were predominantly placed on the forearm, while Productivity icons were preferred in the upper body region. Furthermore, we identified significant associative relationships between icon categories: Leisure and Social Media icons, as well as Productivity and Social Media icons, frequently co-occurred. These patterns underscore the intuitive and potentially synergistic nature of these category pairings within on-body menus.

The insights derived from our studies offer valuable guidance for designers and developers looking to optimize the usability of on-body menus in VR. The evidence supports integrating first-person and mirror perspectives to significantly improve the speed and precision of the menu creation process. However, it is crucial to acknowledge users' strong personal preferences in their choice of perspective during this process. Moreover, the natural propensity to associate specific application categories with certain body locations provides a foundation for intuitive menu design. By understanding and leveraging the associations between application categories, designers can devise more efficient and complementary menu configurations on the body.

\begin{acks}

Xiang Li is supported by the China Scholarship Council (CSC) International Cambridge Scholarship (No. 202208320092). Per Ola Kristensson is supported by the EPSRC (grant EP/W02456X/1). We thank David Lindlbauer (Carnegie Mellon University), Yuzheng Chen (Xi'an Jiaotong-Liverpool University), Xiaohang Tang (Virginia Tech), and Xian Wang (The Hong Kong Polytechnic University) for their feedback and discussions at the early stages of this work. We also extend our gratitude to the anonymous reviewers for their insightful comments, and to our participants for their time.

\end{acks}

\bibliographystyle{ACM-Reference-Format}
\bibliography{onbody}

\appendix

\section{List of Icons}

We categorized 15 icons into five groups: Social Media: Weibo, WeChat, and Tencent QQ; Productivity: Excel, Word, and PowerPoint; Leisure: BiliBili, YouTube, and TikTok; Utilities: Camera, AppStore, and Clock; Other: Google Maps, Ele.me, and Uber.

\begin{figure}[h]
    \centering
    \includegraphics[width=0.3\linewidth]{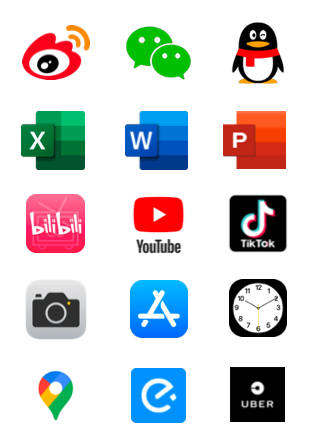}
    \caption{The screenshot of 15 icons that we chose for our studies.}
    \Description{Screenshot of the on-body mapping for 15 virtual icons representing various applications, categorized as: Social Media (Weibo, WeChat, and Tencent QQ), Productivity (Excel, Word, and PowerPoint), Leisure (BiliBili, YouTube, and TikTok), Utilities (Camera, AppStore, and Clock), and Other (Google Maps, Ele.me, and Uber).}
    \label{fig:icons}
\end{figure}








\end{document}